\documentclass[lettersize,journal]{IEEEtran}
\usepackage{amsmath,amsfonts}
\usepackage{algorithm}
\usepackage{graphicx}
\usepackage{multirow}
\usepackage{amsmath}
\usepackage[noend]{algpseudocode}
\usepackage{algpseudocode}
\usepackage{array}
\usepackage[caption=false,font=normalsize,labelfont=sf,textfont=sf]{subfig}
\usepackage{textcomp}
\usepackage{stfloats}
\usepackage{url}
\usepackage{verbatim}
\usepackage{hyperref}
\usepackage{xcolor}
\usepackage{scalerel}
\usepackage{tikz}
\usepackage{balance}

\hypersetup{
	colorlinks,
	citecolor={blue!100!black},
	urlcolor={black!100!black},
	linkcolor={blue!100!black}}

\usetikzlibrary{svg.path}

\definecolor{orcidlogocol}{HTML}{A6CE39}

\tikzset{
	orcidlogo/.pic={
		\fill[orcidlogocol] svg{M256,128c0,70.7-57.3,128-128,128C57.3,256,0,198.7,0,128C0,57.3,57.3,0,128,0C198.7,0,256,57.3,256,128z};
		\fill[white] svg{M86.3,186.2H70.9V79.1h15.4v48.4V186.2z}
		svg{M108.9,79.1h41.6c39.6,0,57,28.3,57,53.6c0,27.5-21.5,53.6-56.8,53.6h-41.8V79.1z M124.3,172.4h24.5c34.9,0,42.9-26.5,42.9-39.7c0-21.5-13.7-39.7-43.7-39.7h-23.7V172.4z}
		svg{M88.7,56.8c0,5.5-4.5,10.1-10.1,10.1c-5.6,0-10.1-4.6-10.1-10.1c0-5.6,4.5-10.1,10.1-10.1C84.2,46.7,88.7,51.3,88.7,56.8z};
	}
}

\newcommand\orcidicon[1]{\href{https://orcid.org/#1}{\mbox{\scalerel*{
				\begin{tikzpicture}[yscale=-1,transform shape]
					\pic{orcidlogo};
				\end{tikzpicture}
			}{|}}}}

\def\BibTeX{{\rm B\kern-.05em{\sc i\kern-.025em b}\kern-.08em
    T\kern-.1667em\lower.7ex\hbox{E}\kern-.125emX}}

\begin{document}
	
\title{Towards Opinion Shaping: A Deep Reinforcement Learning Approach in Bot-User Interactions}

\author{Farbod Siahkali\textsuperscript{\orcidicon{0000-0003-4399-2567}\,}, Saba Samadi\textsuperscript{\orcidicon{0000-0001-7410-8899}\,}, Hamed Kebriaei\textsuperscript{\orcidicon{0000-0002-3781-2163}\,}, \IEEEmembership{Senior Member, IEEE}}

\markboth{IEEE Transactions on Computational Social Systems, Vol.~1 No.~9, June~2024}%
{}

\maketitle

\begin{abstract}
This paper aims to investigate the impact of interference in social network algorithms via user-bot interactions, focusing on the Stochastic Bounded Confidence Model (SBCM). This paper explores two approaches: positioning bots controlled by agents into the network and targeted advertising under various circumstances, operating with an advertising budget. This study integrates the Deep Deterministic Policy Gradient (DDPG) algorithm and its variants to experiment with different Deep Reinforcement Learning (DRL). Finally, experimental results demonstrate that this approach can result in efficient opinion shaping, indicating its potential in deploying advertising resources on social platforms.
\end{abstract}


\section{Introduction}\label{sec1-intro}
\IEEEPARstart{O}{pinion} dynamics studies the evolution and convergence of each individual's opinion on a subject through interaction with people.
Due to the spread of social media platforms, people's opinions are subject to frequent transformations. People are faced with various ideas that can shape their perspectives and decisions at any moment. Opinion dynamics is a crucial field with broad applications, essential for predicting \cite{socio-informed, politic} and controlling \cite{plitjour} final views and polarities. It is used in marketing, business, and finance to forecast customer behavior despite changing trends \cite{fin2}, and in politics to manage opinions based on responses to advertisements and others' perspectives. It also helps control social phenomena and classify opinions \cite{shift}.

Three scenarios are typically studied when analyzing social network dynamics: consensus, polarization, and clustering \cite{threegroups}. Consensus is a state in which everyone in a network converges on the same idea or a similar set of opinions \cite{cons}. Polarization occurs when a society is divided into two subgroups with opposing opinions, often leading to social tension or conflict \cite{pol}. Clustering means that individuals are grouped with the same idea, but each group has a different opinion from the other group \cite{clus}. 
Opinion dynamics is a vast field, and many models have been proposed to simulate this phenomenon and analyze individual interactions.

\subsection{Opinion Dynamics Models}
The simplest and the most classical model is the DeGroot model \cite{DeGroot}. In this model, individuals update their opinions by averaging the opinions of their neighbors. Another model is the Voter model, in which users randomly adopt their neighbor's opinion at each time step \cite{voter}. The Bounded Confidence Model is a widely used model in which individuals update their opinions only based on others' opinions within their range \cite{bdcm}.
In \cite{dwmodel}, the authors present a dynamic opinion formation model known as the Deffuant–Weisbuch (DW) model, initially proposed by \cite{dwmodelbase}. 
Each user \(u \in \{1, \ldots, n\}\) is characterized by a real-valued opinion \(x_u(t) \in \mathbb{R}\) at each time-step \(t\). The confidence bounds of the users, denoted by \(r_u > 0\), are assumed to follow a non-increasing order.
The DW model \cite{dwmodel} exhibits homogeneity if \(r_1 = \ldots = r_n\) and heterogeneity otherwise. This model provides a framework for studying opinion dynamics in groups with varying confidence levels, offering insights into consensus-seeking behavior.

As research progressed, more sophisticated models emerged, such as the Axelrod and Hegselmann-Krause (HK) models, which introduced concepts like cultural dissemination \cite{Axelrod} and bounded confidence \cite{hk}. The HK model \cite{hk} suggests that people only interact if their opinions fall within a specific confidence interval. These models illustrated how individuals might influence each other's opinions, leading to diverse dynamics from consensus to polarization.
The Stochastic Bounded Confidence Model (SBCM) is one of the latest models used to study opinion dynamics \cite{sbcm0}. It incorporates stochastic elements and a bounded confidence mechanism in selecting neighbors, which provides a more realistic representation of how opinions change within social networks \cite{sbcm1, sbcm2}. This model's randomness makes it suitable for investigating opinion dynamics in modern social media environments \cite{socio-informed}.
The update for opinions in this model is expressed as:
\begin{equation}
	x_u(t + 1) = x_u(t) + \frac{\mu}{|\mathcal{N}_u(t)|} \sum_{v \in \mathcal{N}_u(t)} {(x_v(t) - x_u(t))}
	\label{eq:1}
\end{equation}
Here, $\mathcal{N}_u(t)$ represents the set of neighbors for user $u$ at time step $t$. Also, $|\mathcal{N}_u(t)|$ denotes the number of neighbors. The outcome of these interactions can result in different public opinion formations.
The main part of SBCM is that these neighbors are sampled according to a probability distribution. The sampling process is formalized as follows: the probability \( p(z_{uv}^{t} = 1) \) of individual \( u \) engaging with individual \( v \) at time \( t \) is proportional to the inverse of the opinion distance raised to a power \( \epsilon \), as shown in equation:

\begin{equation}
	p(z_{uv}^{t} = 1) = \frac{\left| x_u(t) - x_v(t) \right|^{-\epsilon}}{\sum_{v'} \left| x_u(t) - x_{v'}(t) \right|^{-\epsilon}}
 \label{eqSBCM}
\end{equation}
Here, \( \epsilon > 0 \) suggests that individuals with closely aligned opinions are more inclined to interact and exert mutual influence, whereas \( \epsilon < 0 \) indicates that individuals are more likely to interact with those holding divergent views.

\subsection{Opinion Shaping Algorithms}

Despite extensive research \cite{leader, consop}, opinion shaping is still a challenging task. The paper \cite{shape} explores a modified DeGroot model \cite{DeGroot} for opinion propagation in social networks. It introduces a control parameter affecting a subset of agents, including stubborn and uncontrollable agents. Theoretical analysis and numerical experiments are conducted, including situations where only specific agent interactions are observable. In the \cite{opbots}, a variant of the voter model \cite{voterold} was introduced to analyze opinion dynamics in networks with both individuals and bots. A noteworthy aspect was distinguishing the influence of bots from that of individuals.

The paper \cite{opadv} addresses the influence of external factors on group opinion evolution in social networks. The study examines how advertising affects networks with rebels using a novel DeGroot model \cite{DeGroot}. It proposes an effective advertising strategy to enhance support in both structurally balanced and unbalanced networks. Additionally, the paper \cite{opop} introduces a framework for analyzing the controllability of opinion dynamics within social networks. The study demonstrates the potential for controlling opinions in a population by utilizing a committed node that consistently advocates the opposing opinion and remains immune to external influence. 

Opinion shaping has been approached through diverse perspectives \cite{shape}. Various methods and algorithms have been developed to simulate real-world scenarios. These algorithms use reinforcement learning as their strategy to control \cite{csh0} and shape opinions \cite{csh1}.
Traditional methods have been used for persuasion, including subtle influence like information filtering \cite{persuasion}. In the digital age, artificial intelligence (AI) has opened new frontiers, providing tools that can analyze and influence public sentiment.

In this context, our study proposes an innovative approach to perform opinion shaping within the SBCM \cite{socio-informed, sbcm0}. The DDPG algorithm \cite{ddpg} is integrated to explore how targeted strategies can be developed and implemented to shape opinions effectively in social networks. By combining the DDPG \cite{ddpg} with the simulation environment of SBCM \cite{socio-informed, sbcm0}, this research aims to provide new insights into the potential and challenges of AI-driven opinion-shaping in social platforms. 
The main contributions of this article can be summarized as follows:

\begin{enumerate}
	\item We incorporated the user-bot interaction scenario into the SBCM \cite{socio-informed, sbcm0}, employing a singular agent with comprehensive observation through the DDPG algorithm \cite{ddpg}.
	
	\item Our research also investigates targeted advertising under a limited budget. In this scenario, the agent indicates a location with an advertising range to shift users' opinions to the desired opinion.

        \item We experimented both scenarios with diverse configurations to identify the optimal approach for each setup.
\end{enumerate}

\section{Proposed Method}\label{sec3-meth}

This section will explore two distinct methods to perform opinion shaping. Firstly, bots controlled by agents will be injected into the network. Secondly, targeted advertising will be implemented in a specific area under an advertising budget. These approaches are more feasible and practical compared to other methods discussed in Section \ref{sec1-intro}. Here, the DDPG algorithm \cite{ddpg} has been utilized for training. It is designed to handle environments with continuous action spaces. By combining the strengths of deterministic policy gradients with deep learning, DDPG effectively learns optimal policies in complex, high-dimensional settings. In addition, the convergence analysis of Deterministic Policy Gradient (DPG) has been discussed in \cite{DPG}.

\subsection{Interaction Between Bots and Users in SBCM}

\begin{algorithm}[t]
	\caption{User-Bot Interaction in SBCM}
	\begin{algorithmic}[1]
		\Procedure{DDPG\_Train}{N, T}
		\State Initialize Actor: $\pi(s;\theta^\pi)$ and $\pi'(s;\theta^{\pi'})$
		\State Initialize Critic: $Q(s, a;w^{Q})$ and $Q'(s, a;w^{Q'})$
		\State Initialize Replay Buffer $\mathcal{R}$
		\State Initialize noise process $\mathcal{N}$
		\For {episode = 1 to M}
		\State Reset noise process: $\mathcal{N}.\mathrm{reset}()$
		\State Sample initial user opinions: $u_0 \sim U(-1, 1)$
		\For {t = 1 to T}
		\State Augment state with time: $s_{t-1} = [u_{t-1}, t]$
		\State Select action: $a_{t-1} = \pi(s_{t-1}) + \mathcal{N}.\mathrm{sample}()$
		\State $u_t, r_t \leftarrow \mathrm{SBCM}(N, u_{t-1}, a_{t-1})$
		\State Store $(s_{t-1}, a_{t-1}, r_t, s_{t})$ in $\mathcal{R}$
		\If {size of $\mathcal{R}$ $\geq$ batch\_size}
		\State Update $Q$ and $\pi$ using mini-batch from $\mathcal{R}$
		\State Softly update $\pi'$ and $Q'$ with $\tau$
		\EndIf
		\EndFor
		\EndFor
		\EndProcedure
	\end{algorithmic}
	\label{ddpgsbcm_bot}
\end{algorithm}

In Algorithm \ref{ddpgsbcm_bot}, the DDPG method \cite{ddpg} is utilized to optimize interactions between users and bots within the SBCM framework \cite{socio-informed, sbcm0}. This algorithm makes use of an actor denoted as $\pi$ and a critic denoted as $Q$, with parameters $\theta^\pi$ and $w^Q$ respectively. All experiences are stored in a replay buffer $R$, which is then used to update the actor and critic. Additionally, a noise process $N$ is added to the actor's actions to ensure exploration in the environment. Initial users' opinions $u_0$ is from a uniform distribution $U(-1,1)$
In each time step of an episode, which consists of $T$ time-steps, the actor, $\pi$ with parameters $\theta^\pi$, sets the bots' opinions based on the input state representation, which is a combination of the users' opinions vector $u_t$ and the encoded time step $t$. The bot-user SBCM environment is then simulated for one iteration, which repeats until the end of the episode. The actor-critic policy of DDPG \cite{ddpg} is updated based on the calculated reward.
In this bot-user interaction setup, reward at each time-step \(r_t\) is defined as the change in the average opinion, modulated by the time-step:
\begin{equation}
	r_t = (\frac{t}{T})\times [{\mu}(u_t) - {\mu}(u_{t-1})]
\end{equation}
where ${\mu}(u_t)$ represents the mean opinion at time $t$. 
The term \( \frac{t}{T} \) is a scaling factor for the reward to increase over time.

Hence, the purpose of the DDPG \cite{ddpg} is to maximize users' opinion shift. The algorithm learns that shifting opinions becomes more critical near the end of an episode, resulting in higher rewards. 
The main challenge is that, according to the bounded confidence models, extreme opinions will result in either an empty set of neighbors or a near-zero probability of interaction with users. Therefore, the actor needs to set the bots' opinions in a way that increases the likelihood of interaction with users and also aims to shift their opinions to the desired state.
Integrating DRL techniques with SBCM \cite{socio-informed, sbcm0} creates a dynamic method of shaping opinions in user-bot interactions. The Actor-Critic architecture used in DDPG \cite{ddpg} helps to learn efficient strategies for shaping opinions.

\subsection{Advertising In SBCM}

\begin{algorithm}[t]
	\caption{Advertising in SBCM}
	\begin{algorithmic}[1]
		\Procedure{DDPG\_Train}{N, T, $B_0$}
		\State Initialize Actor: $\pi(s;\theta^\pi)$ and $\pi'(s;\theta^{\pi'})$
		\State Initialize Critic: $Q(s, a;w^{Q})$ and $Q'(s, a;w^{Q'})$
		\State Initialize Replay Buffer $\mathcal{R}$
		\State Initialize noise process $\mathcal{N}$
		\For{episode = 1 to M}
		\State Reset noise process: $\mathcal{N}.\mathrm{reset}()$
		\State Sample initial user opinions: $u_0 \sim U(-1, 1)$
		\State Initialize remaining budget: $B = B_0$
		\For{t = 1 to T}
		\State State Representation: $s_{t-1} = [u_{t-1}, t, B]$
		\State Select action: $a_{t-1} = \pi(s_{t-1}) + \mathcal{N}.\mathrm{sample}()$
		\State $(u_{t-1}^\prime, C_t) \leftarrow \text{Advertisement}(u_{t-1}, a_{t-1})$
		\State $B \leftarrow B - C_t$
		\State $u_t, r_t \leftarrow \mathrm{SBCM}(N, u_{t-1}^\prime, a_{t-1})$
		\If{$B < 0$}
		\State $r_t \leftarrow r_t - C_t$
		\EndIf
		\State Store $(s_{t-1}, a_{t-1}, r_t, u_t, B)$ in $\mathcal{R}$
		\If{size of $\mathcal{R}$ $\geq$ batch\_size}
		\State Update $Q$ and $\pi$ using mini-batch from $\mathcal{R}$ 
		\State Softly update $\pi'$ and $Q'$ with $\tau$
		\EndIf
		\EndFor
		\EndFor
		\EndProcedure
	\end{algorithmic}
	\label{ddpgsbcm_adv}
\end{algorithm}

In the advertising scenario, the dynamics of user opinions within SBCM \cite{socio-informed, sbcm0} are being investigated when external stimuli, such as targeted advertising, are introduced. The focus is on understanding how the opinion state is affected by advertising, characterized by the opinion of its content and the spread of its influence.
Algorithm \ref{ddpgsbcm_adv} explains how the advertising scenario is integrated into the simulation. At each time step $t$, the actor $\pi$ receives a state representation that includes users' opinions $u_t$, encoded time-step $t$, and remaining budget $B$. It outputs two actions: advertising location $A_l \in [-1,1]$ and advertising range $A_r \in [0,1]$. After that, the SBCM \cite{socio-informed, sbcm0} simulation runs the opinion dynamics and advertising process for one time-step. 
The advertising process is similar to the SBCM process described in equation \ref{eqSBCM}, only with the difference that the selected user $u$ will interact with the advertising opinion location $A_l$ instead of interacting with $v$. In other words, it is another SBCM simulation only for a subset of users in the advertising range with a fixed $v$ to interact with.
Finally, the reward function, taking into account the budget's state, can be formulated as follows:
\begin{equation}
r_t = 
\begin{cases} 
	(\frac{t}{T}) \times [{\mu}(u_t) - {\mu}(u_{t-1})], & \text{if } B_t \geq 0 \\
	(\frac{t}{T}) \times [{\mu}(u_t) - {\mu}(u_{t-1})] - C_t, & \text{if } B_t < 0
\end{cases} 
\end{equation}
Where \( C_t \) represents the cost of advertising and \( B_t \) is the remaining budget at time-step \(t\) after the advertising cost is applied.
The terms \( C_r \) and \( C_o \) are scaling factors. The cost of advertising and opinion shift is determined by exponential functions, with $p$ and $q$ representing the exponents scaling the spread and opinion shift, respectively. 
If the actor runs out of budget before the end of an episode, it has the option to continue, but any extra costs $C_t$ will be taken out of the reward $r_t$. This encourages the actor to budget carefully from the beginning.
\begin{equation}
\begin{aligned}
	{Cost_{\text{range}}} &= C_r \times \left(e^{(A_r \times p)} - 1\right) \\
	{Cost_{\text{opinion}}} &= C_o \times \left(e^{(\lvert A_l \rvert \times q)} - 1\right) \\
	{C_t} &= {Cost_{\text{range}}} + {Cost_{\text{opinion}}}
\end{aligned}
\label{eq_cost_adv}
\end{equation}
These exponential equations were selected to represent a realistic non-linear increase in cost based on the advertising range and the absolute value of the advertising location, as advertising in very isolated locations costs more. Ultimately, the total cost is the sum of the costs for the advertising range and location.

The DDPG algorithm \cite{ddpg} is used to optimize the location and range of advertisements, considering a budget and diminishing returns on investment. The advertising strategy is continuously improved through the actor-critic approach of DDPG \cite{ddpg}, where the actor network proposes advertising actions and the critic network evaluates their effectiveness. The algorithm aims to maximize opinion shift toward the desired stance while respecting budget and user responsiveness constraints across multiple simulated episodes.

These two approaches discussed in this section will be experimented with by comparing different settings and configurations in the SBCM \cite{socio-informed, sbcm0}. The characteristics of the final state of users' opinions will be reported to validate the practicality of our proposed opinion shaping approaches.

\begin{figure}[b]
	\centering
	\includegraphics[width=\linewidth]{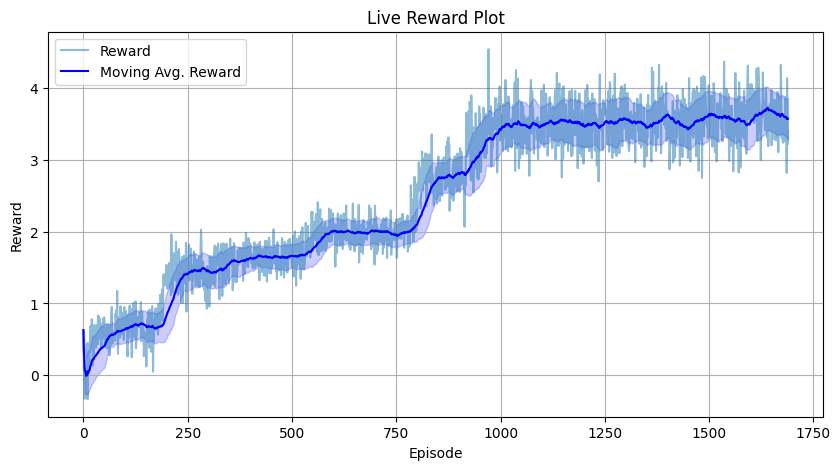}
	\caption{Average reward of the actor over episodes. This plot provides insights into how the agent's performance evolves over training episodes.}
	\label{fig:reward_plot}
\end{figure}

\section{Experiments}\label{sec4-work}

This section evaluates the effectiveness of the proposed methods for shaping opinions in two distinct consensus scenarios. The experiments focus on the impact of these methods by varying the number of bots and budget parameters. In all analyses, we study a social network of 200 users with 200 time-steps in each episode, and an Ornstein-Uhlenbeck process models the noise in both algorithms.

\subsection{Evaluation of User-Bot Interaction Scenario in SBCM}

Both scenarios has been trained for over 1700 episodes and Fig. \ref{fig:reward_plot} illustrates the average reward obtained by the actor in bot-user scenario, indicating the actor’s learning process and enhanced strategy for influencing user opinions within the SBCM \cite{sbcm0}. The figure also indicates the actor's growing proficiency in optimizing user-bot interactions for opinion shaping.

\begin{figure}[t]
	\centering
	\includegraphics[width=\linewidth]{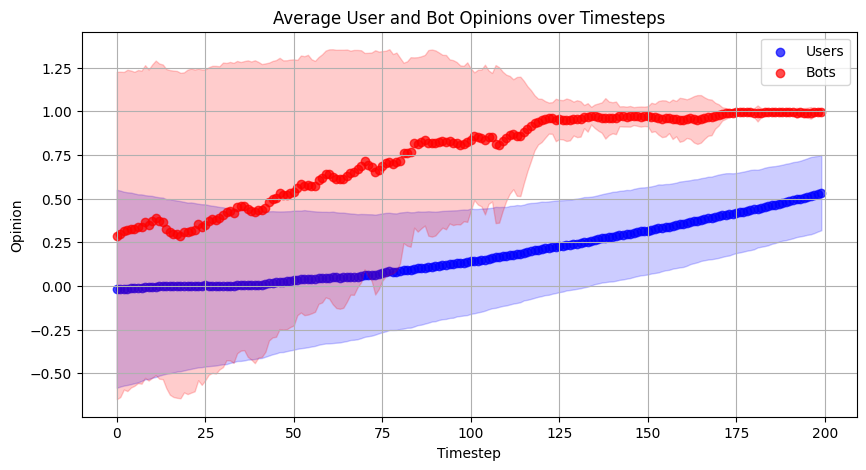}
	\caption{Visualization of user and bot opinions over time steps. The plot demonstrates the dynamics of opinions and the influence of bots.}
	\label{fig:opinion_plot}
\end{figure}

Fig. \ref{fig:opinion_plot} illustrates the interaction between users and bots within the SBCM \cite{socio-informed, sbcm0} in an episode. It shows how the opinions of users and bots evolve and influence each other throughout the simulation.
The bots' opinions, controlled by agents, have learned how to position their opinions in order to maximize opinion shifting in users. The agents have to learn how to effectively position the bots' opinions within the bounds of users' opinions to increase the chance of interaction. This can be seen in Fig. \ref{fig:opinion_plot}, as the bots' opinions are closer to the users' opinion mean, then gradually increasing towards the desired opinion state (which here is 1).

\begin{table}[t]
	\centering
	\caption{Evaluating mean and std of opinions in the user-bot scenario for 1000 episodes and the varying number of bots, \(\mu\), and \(\epsilon\).}
	\begin{tabular}{ccccc}
		\hline
		\multirow{2}{*}{Number of Bots} & \multicolumn{2}{c}{\(\mu = 0.1, \epsilon = -2.0\)} & \multicolumn{2}{c}{\(\mu = 0.05, \epsilon = -1.0\)} \\  
		&    Mean    &  STD       &     Mean  &   STD     \\ \hline
		
		20      &   0.556    & 0.071      &    0.573  &   0.074     \\
		15      &   0.481    & 0.075      &    0.117  &   0.074     \\
		10      &   0.389    & 0.083      &    0.074  &   0.072      \\ \hline
	\end{tabular}
	\label{tab:user_bot_res}
\end{table} 

To concretely illustrate the impact of bot intervention on user opinions, Table \ref{tab:user_bot_res} presents a detailed analysis of the mean and standard deviation (STD) of user opinions after 1000 episodes across various configurations of bot numbers, $\mu$, and $\epsilon$ parameters (refer to equations \ref{eq:1}, \ref{eqSBCM}). This table is pivotal in understanding the subtle influence of bots on opinion formation within social networks. As the number of bots increases, there is a noticeable shift in the average opinion, indicating the effectiveness of bots in influencing user perspectives. Additionally, the varying STDs reflect the diversity in final opinion state of users.

\subsection{Evaluation of Advertising Scenario in SBCM}

In this section, we present the experiments and results of the second proposed method for opinion shaping, called the advertising scenario. To proceed with the simulation, the parameters in equation \ref{eq_cost_adv} are assigned the following values: $C_r$ and $C_o$ are set to 0.05, while $p$ and $q$ are both set to 1.5.

\begin{figure}[t]
	\centering
	\includegraphics[width=\linewidth]{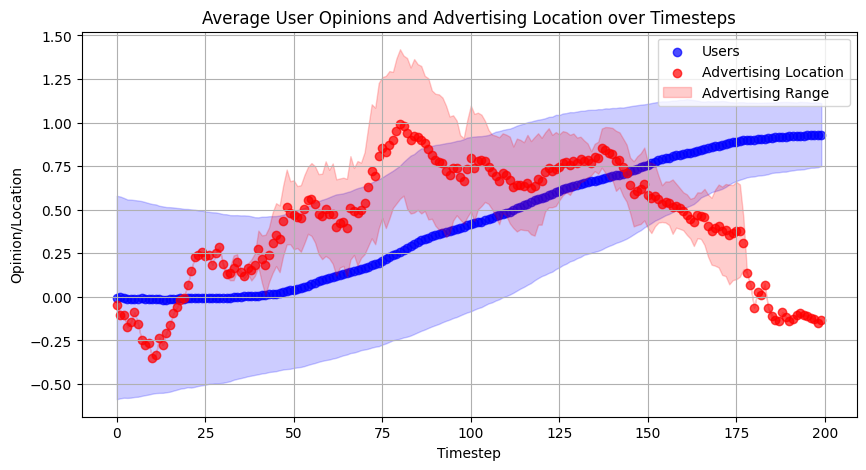} 
	\caption{The trajectory of user opinions and the influence of targeted advertising in a single episode with $\mu=0.1$, and $\epsilon$ $= -2.0$.}
	\label{fig:adv}
\end{figure}

\begin{table}[t]
	\centering
	\caption{Evaluating mean and std of opinions in advertising scenario for 1000 episodes and varying budget amount, \(\mu\), and \(\epsilon\).}
	\begin{tabular}{ccccc}
		\hline
		\multirow{2}{*}{Budget} & \multicolumn{2}{c}{\(\mu = 0.1, \epsilon = -2.0\)} & \multicolumn{2}{c}{\(\mu = 0.05, \epsilon = -1.0\)} \\ 
		&  Mean    &   STD    & Mean    & STD   \\ \hline 
		20  &  0.851   &  0.190   & 0.321   & 0.433  \\ 
		10  &  0.735   &  0.214   & 0.267   & 0.423   \\ 
		5   &  0.617   &  0.224   & 0.226   & 0.414   \\ \hline
	\end{tabular}
	\label{tab:adv_res}
\end{table} 

In Fig. \ref{fig:adv}, a simulation of an advertising scenario is depicted. In each episode, the actor determines the advertising location and range, which are represented by the red line. The actor operates within a budget and aims to spend it effectively, as the penalty for exceeding the budget is substantial.
In Fig. \ref{fig:adv}, time-step 175 of the episode indicates a pivotal moment in the actor's learned strategy within the simulation. As the budget approaches depletion, the actor adopts a conservative posture, converging on a strategy that minimizes expenditure. By targeting the neutral opinion (0) with zero advertising range, the actor effectively reduces the scope and advertising costs to the lowest possible value. 
This maneuver reflects an understanding by the actor that, with limited resources, it is crucial to avoid further influence attempts that would incur additional costs. The transition to this conservative approach directly responds to the constraints imposed by the budget (refer to equation \ref{eq_cost_adv}).

Table \ref{tab:adv_res} analyzes different configurations of two consensus settings with varying budgets. It is noticeable that in all scenarios, there has been a shift in the overall opinion state. As the values of $\mu$ and $\epsilon$ moved from the consensus scene towards clustering scenarios (i.e., approaching a value of 0), their influence on users' opinions decreased. 
As we analyze extreme clustering scenes, the standard deviation of users' opinions grows as the number of subgroups increases. This occurs because users are more likely to engage with individuals who share a similar opinion. Therefore, this suggests that a larger budget is required to influence final users' opinions in such situations.

\section{Conclusion}\label{sec5-conc}

In this paper, opinion-shaping in SBCM using the DDPG algorithm was proposed. The main innovation involved introducing agents controlling a group of bots within the social network. The bots, which can set their opinions at will and are immune to users' opinions, were used to steer the average opinion of users toward a desired state. Another proposed method was utilizing an advertiser agent acting under a budget. These agents learned optimal strategies for opinion shaping based on observing users' opinions. The experimental results verify the proposed algorithms' effectiveness and scalability. Important considerations raised by this study are the ethical use of AI in social networks and social media platforms.

Possible future research could investigate other scenarios and more sophisticated networks, comprising multiple sub-networks, each following a different opinion dynamics model. Additionally, adversarial scenarios can be experimented with, where one agent attempts to shift opinions using bots while the other works to identify bots from users within the network.

\bibliographystyle{IEEEtran}
\bibliography{refs}

\begin{thebibliography}{10}
\providecommand{\url}[1]{#1}
\csname url@samestyle\endcsname
\providecommand{\newblock}{\relax}
\providecommand{\bibinfo}[2]{#2}
\providecommand{\BIBentrySTDinterwordspacing}{\spaceskip=0pt\relax}
\providecommand{\BIBentryALTinterwordstretchfactor}{4}
\providecommand{\BIBentryALTinterwordspacing}{\spaceskip=\fontdimen2\font plus
\BIBentryALTinterwordstretchfactor\fontdimen3\font minus
  \fontdimen4\font\relax}
\providecommand{\BIBforeignlanguage}[2]{{%
\expandafter\ifx\csname l@#1\endcsname\relax
\typeout{** WARNING: IEEEtran.bst: No hyphenation pattern has been}%
\typeout{** loaded for the language `#1'. Using the pattern for}%
\typeout{** the default language instead.}%
\else
\language=\csname l@#1\endcsname
\fi
#2}}
\providecommand{\BIBdecl}{\relax}
\BIBdecl

\bibitem{socio-informed}
\BIBentryALTinterwordspacing
M.~Okawa and T.~Iwata, ``Predicting opinion dynamics via
  sociologically-informed neural networks,'' in \emph{Proceedings of the 28th
  ACM SIGKDD Conference on Knowledge Discovery and Data Mining}.\hskip 1em plus
  0.5em minus 0.4em\relax ACM, Aug. 2022. [Online]. Available:
  \url{http://dx.doi.org/10.1145/3534678.3539228}
\BIBentrySTDinterwordspacing

\bibitem{politic}
\BIBentryALTinterwordspacing
E.~Ben-Naim, ``Opinion dynamics: Rise and fall of political parties,''
  \emph{Europhysics Letters (EPL)}, vol.~69, no.~5, p. 671–677, Mar. 2005.
  [Online]. Available: \url{http://dx.doi.org/10.1209/epl/i2004-10421-1}
\BIBentrySTDinterwordspacing

\bibitem{plitjour}
E.~Holder and C.~X. Bearfield, ``Polarizing political polls: How visualization
  design choices can shape public opinion and increase political
  polarization,'' \emph{IEEE Transactions on Visualization and Computer
  Graphics}, vol.~30, no.~1, pp. 1446--1456, 2024.

\bibitem{fin2}
\BIBentryALTinterwordspacing
Q.~Zha, G.~Kou, H.~Zhang, H.~Liang, X.~Chen, C.-C. Li, and Y.~Dong, ``Opinion
  dynamics in finance and business: a literature review and research
  opportunities,'' \emph{Financial Innovation}, vol.~6, no.~1, p.~44, Jan 2021.
  [Online]. Available: \url{https://doi.org/10.1186/s40854-020-00211-3}
\BIBentrySTDinterwordspacing

\bibitem{shift}
Y.~Yi, T.~Castiglia, and S.~Patterson, ``Shifting opinions in a social network
  through leader selection,'' \emph{IEEE Transactions on Control of Network
  Systems}, vol.~8, no.~3, pp. 1116--1127, 2021.

\bibitem{threegroups}
L.~Li, A.~Scaglione, A.~Swami, and Q.~Zhao, ``Consensus, polarization and
  clustering of opinions in social networks,'' \emph{IEEE Journal on Selected
  Areas in Communications}, vol.~31, no.~6, pp. 1072--1083, 2013.

\bibitem{cons}
Y.~Zhang, X.~Chen, Z.~Huang, X.~Li, and Y.~Du, ``A global opinion-influencing
  consensus model based on the degroot,'' in \emph{2022 IEEE 21st International
  Conference on Ubiquitous Computing and Communications
  (IUCC/CIT/DSCI/SmartCNS)}, 2022, pp. 191--197.

\bibitem{pol}
S.~Das, S.~Dutta, S.~Chakraborty, and S.~Biswas, ``Influence of opinion
  formation on polarization in social networks,'' in \emph{2023 14th
  International Conference on Computing Communication and Networking
  Technologies (ICCCNT)}, 2023, pp. 1--7.

\bibitem{clus}
H.~Zhang, J.~Dong, Y.~Zhao, and J.~Hu, ``Opinion formation over clustered
  social networks with intermittent communication,'' in \emph{2023 35th Chinese
  Control and Decision Conference (CCDC)}, 2023, pp. 5018--5023.

\bibitem{DeGroot}
\BIBentryALTinterwordspacing
M.~H. DeGroot, ``Reaching a consensus,'' \emph{Journal of the American
  Statistical Association}, vol.~69, no. 345, pp. 118--121, 1974. [Online].
  Available: \url{http://www.jstor.org/stable/2285509}
\BIBentrySTDinterwordspacing

\bibitem{voter}
E.~Yildiz, A.~Ozdaglar, D.~Acemoglu, and A.~Scaglione, ``The voter model with
  stubborn agents extended abstract,'' in \emph{2010 48th Annual Allerton
  Conference on Communication, Control, and Computing (Allerton)}, 2010, pp.
  1179--1181.

\bibitem{bdcm}
\BIBentryALTinterwordspacing
F.~Ceragioli and P.~Frasca, ``Continuous and discontinuous opinion dynamics
  with bounded confidence,'' \emph{Nonlinear Analysis: Real World
  Applications}, vol.~13, no.~3, pp. 1239--1251, 2012. [Online]. Available:
  \url{https://www.sciencedirect.com/science/article/pii/S1468121811002902}
\BIBentrySTDinterwordspacing

\bibitem{dwmodel}
\BIBentryALTinterwordspacing
G.~Chen, W.~Su, W.~Mei, and F.~Bullo, ``Convergence properties of the
  heterogeneous deffuant–weisbuch model,'' \emph{Automatica}, vol. 114, p.
  108825, 2020. [Online]. Available:
  \url{https://www.sciencedirect.com/science/article/pii/S0005109820300236}
\BIBentrySTDinterwordspacing

\bibitem{dwmodelbase}
\BIBentryALTinterwordspacing
G.~Deffuant, D.~Neau, F.~Amblard, and G.~Weisbuch, ``Mixing beliefs among
  interacting agents,'' \emph{Advances in Complex Systems}, vol.~03, no. 01n04,
  pp. 87--98, 2000. [Online]. Available:
  \url{https://doi.org/10.1142/S0219525900000078}
\BIBentrySTDinterwordspacing

\bibitem{Axelrod}
\BIBentryALTinterwordspacing
R.~Axelrod, ``The dissemination of culture: A model with local convergence and
  global polarization,'' \emph{The Journal of Conflict Resolution}, vol.~41,
  no.~2, pp. 203--226, 1997. [Online]. Available:
  \url{http://www.jstor.org/stable/174371}
\BIBentrySTDinterwordspacing

\bibitem{hk}
\BIBentryALTinterwordspacing
R.~Hegselmann and U.~Krause, ``{Opinion Dynamics and Bounded Confidence Models,
  Analysis and Simulation},'' \emph{Journal of Artificial Societies and Social
  Simulation}, vol.~5, no.~3, pp. 1--2, 2002. [Online]. Available:
  \url{https://ideas.repec.org/a/jas/jasssj/2002-5-2.html}
\BIBentrySTDinterwordspacing

\bibitem{sbcm0}
F.~Baccelli, A.~Chatterjee, and S.~Vishwanath, ``Stochastic bounded confidence
  opinion dynamics,'' in \emph{53rd IEEE Conference on Decision and Control},
  2014, pp. 3408--3413.

\bibitem{sbcm1}
Q.~Liu and X.~Wang, ``Social learning with bounded confidence and probabilistic
  neighbors,'' in \emph{2013 IEEE International Symposium on Circuits and
  Systems (ISCAS)}, 2013, pp. 2303--2306.

\bibitem{sbcm2}
\BIBentryALTinterwordspacing
F.~Baumann, P.~Lorenz-Spreen, I.~M. Sokolov, and M.~Starnini, ``Emergence of
  polarized ideological opinions in multidimensional topic spaces,''
  \emph{Phys. Rev. X}, vol.~11, p. 011012, Jan 2021. [Online]. Available:
  \url{https://link.aps.org/doi/10.1103/PhysRevX.11.011012}
\BIBentrySTDinterwordspacing

\bibitem{leader}
Z.~Zhao, L.~Shi, T.~Li, J.~Shao, and Y.~Cheng, ``Opinion dynamics of social
  networks with intermittent-influence leaders,'' \emph{IEEE Transactions on
  Computational Social Systems}, vol.~10, no.~3, pp. 1073--1082, 2023.

\bibitem{consop}
C.~Ge, Y.~Liu, and X.~Chen, ``Group opinion consensus under moderate
  guidance,'' \emph{IEEE Transactions on Computational Social Systems}, pp.
  1--8, 2023.

\bibitem{shape}
V.~S. Borkar and A.~Reiffers-Masson, ``Opinion shaping in social networks using
  reinforcement learning,'' \emph{IEEE Transactions on Control of Network
  Systems}, vol.~9, no.~3, pp. 1305--1316, 2022.

\bibitem{opbots}
A.~Shukla, N.~Sahasrabudhe, and S.~Moharir, ``Opinion dynamics in the presence
  of bots,'' in \emph{2022 IEEE International Conference on Signal Processing
  and Communications (SPCOM)}, 2022, pp. 1--5.

\bibitem{voterold}
\BIBentryALTinterwordspacing
R.~A. Holley and T.~M. Liggett, ``{Ergodic Theorems for Weakly Interacting
  Infinite Systems and the Voter Model},'' \emph{The Annals of Probability},
  vol.~3, no.~4, pp. 643 -- 663, 1975. [Online]. Available:
  \url{https://doi.org/10.1214/aop/1176996306}
\BIBentrySTDinterwordspacing

\bibitem{opadv}
M.-H. Yang, J.-W. Yi, and L.~Chai, ``Opinion dynamics of the degroot model with
  rebels and advertising,'' in \emph{2021 China Automation Congress (CAC)},
  2021, pp. 7493--7498.

\bibitem{opop}
\BIBentryALTinterwordspacing
Z.~Liu, J.~Ma, Y.~Zeng, L.~Yang, Q.~Huang, and H.~Wu, ``On the control of
  opinion dynamics in social networks,'' \emph{Physica A: Statistical Mechanics
  and its Applications}, vol. 409, pp. 183--198, 2014. [Online]. Available:
  \url{https://www.sciencedirect.com/science/article/pii/S0378437114003550}
\BIBentrySTDinterwordspacing

\bibitem{csh0}
C.~Ancona, P.~De~Lellis, and F.~Lo~Iudice, ``Influencing opinions in a
  nonlinear pinning control model,'' \emph{IEEE Control Systems Letters},
  vol.~7, pp. 1945--1950, 2023.

\bibitem{csh1}
F.~Dietrich, S.~Martin, and M.~Jungers, ``Control via leadership of opinion
  dynamics with state and time-dependent interactions,'' \emph{IEEE
  Transactions on Automatic Control}, vol.~63, no.~4, pp. 1200--1207, 2018.

\bibitem{persuasion}
\BIBentryALTinterwordspacing
H.~Hu and N.~Xu, ``Persuasion process on social networks,'' \emph{International
  Journal of Modern Physics C}, vol.~33, no.~04, p. 2250052, 2022. [Online].
  Available: \url{https://doi.org/10.1142/S0129183122500528}
\BIBentrySTDinterwordspacing

\bibitem{ddpg}
T.~P. Lillicrap, J.~J. Hunt, A.~Pritzel, N.~Heess, T.~Erez, Y.~Tassa,
  D.~Silver, and D.~Wierstra, ``Continuous control with deep reinforcement
  learning,'' 2019.

\bibitem{DPG}
H.~Xiong, T.~Xu, L.~Zhao, Y.~Liang, and W.~Zhang, ``Deterministic policy
  gradient: Convergence analysis,'' in \emph{Uncertainty in Artificial
  Intelligence}.\hskip 1em plus 0.5em minus 0.4em\relax PMLR, 2022, pp.
  2159--2169.

\end{thebibliography}

\end{document}